\documentclass[aps,pra,superscriptaddress,floatfix,showpacs,10pt, twocolumn]{revtex4}
\usepackage[dvips]{graphicx}
\usepackage{amsmath,times}

\begin{document}

\title{Implementation of the Grover search algorithm with Josephson charge qubits }
\author{Xiao-Hu Zheng\footnote{Electronic address:
xhzheng@ahu.edu.cn}, Ping Dong, Zheng-Yuan Xue and Zhuo-Liang
Cao\footnote{Electronic address: zlcao@ahu.edu.cn (Corresponding
Author)}}

\affiliation{Key Laboratory of Opto-electronic Information
Acquisition and Manipulation, Ministry of Education, School of
Physics and Material Science, Anhui University, Hefei, 230039, P R
China}

\begin{abstract}
A scheme of implementing the Grover search algorithm based on
Josephson charge qubits has been proposed, which would be a key step
to scale more complex quantum algorithms and very important for
constructing a real quantum computer via Josephson charge qubits.
The present scheme is simple but fairly efficient, and easily
manipulated because any two-charge-qubit can be selectively and
effectively coupled by a common inductance. More manipulations can
be carried out before decoherence sets in. Our scheme can be
realized within the current technology.

\end{abstract}

\maketitle

It is well anticipated that a quantum computer could accomplish a
huge task, which a classical computer may never fulfill in an
acceptable time. As a result, the realization of the quantum
computer has attracted not only many physicists but also many
scientific researchers such as computer scientists, electrical
engineers and so on. The implementation of quantum algorithms is the
basis of inventing quantum computers. Two classes of quantum
algorithms have shown great promise of the quantum computer. One is
based on Shor's Fourier transform including quantum factoring
\cite{Shor} and Deutsh-Jozsa algorithm \cite{D}. The other is based
on the Grover quantum search algorithm \cite{Grover}, which is
quadratic speedup compared with the classical one. The Grover search
algorithm is very important because many techniques based on the
search algorithm are used universally in our lives. The Grover
search algorithm is fairly efficient in looking for one item in an
unsorted database of size $N\equiv2^{n}$ \cite{Grover,H}. In order
to achieve the task, classical algorithm needs $O(N)$ queries but
$O(\sqrt{N})$ queries by the Grover search algorithm. The efficiency
of the algorithm has been manipulated experimentally in few-qubit
cases via Nuclear Magnetic Resonance (NMR) \cite{1,2} and optics
\cite{3,4}.

The Grover search algorithm can be used to search one item from
$2^{n}$ items with $n$ data qubits and one auxiliary working qubit.
The process can be concluded as the following four steps: Firstly,
prepare the $n+1$ qubits, in which $n$ data qubits are in
$|0\rangle^{\otimes n}$, and one auxiliary working qubit in
$|1\rangle_{n+1}$. And perform the $n+1$ Hadamard transform on the
$n+1$ qubits. Secondly, apply the oracle. The auxiliary working
qubit can be omitted after the second step. Thirdly, perform the $n$
Hadamard transform on the $n$ data qubits, and apply a phase shift
to the data qubits except $|0\rangle^{\otimes n}$, which can be
described by the unitary operator $2|0\rangle^{\otimes
n}\langle0|-I$, where $I$ is identity operation on the data qubits.
And perform the $n$ Hadamard transformations on the $n$ data qubits
again. Finally, repeat Steps $2\rightarrow 3$ with a finite number
of times until obtaining a solution to the search problem with high
probability. And measure the $n$ data qubits. The number of
repetitions \cite{M} for obtaining a finite item is $R=CI
(\frac{arccos\sqrt{1/N}}{2arccos \sqrt{N-1/N}})$, which is bounded
above by $\pi \sqrt{N}/4$.

Because of several advantages over other qubits, e.g., having
large-scale integration, relatively high quantum coherence, being
manipulated more easily, and being manufactured by modern
microfabrication techniques, Josephson charge \cite{21-1,21-2, 22}
and phase \cite{23Mooij, 24} qubits, based on the macroscopic
quantum effects in low-capacitance Josephson junction circuits
\cite{Makhlin,You2005}, have recently been used in quantum
information processing. Some striking experimental phenomena
\cite{Nakamura,van} have shown that the Josephson charge and phase
qubits are promising candidates of solid-state qubits in quantum
information processing. In particular, recent experimental
realization of a single charge qubit demonstrates that it is hopeful
to construct the quantum computer by means of Josephson charge
qubits \cite{Nakamura2}. Accordingly, implementation of quantum
algorithm by Josephson charge qubits is of great importance.
Recently, many schemes of quantum algorithms with Josephson charge
qubits have been proposed. For example, Vartiainen \emph{et al.}
\cite{Vartiainen} have implemented a Shor's factorization algorithm
on a Josephson charge qubit register. Fazio \emph{et al.}
\cite{Fazio} have realized a simple solid-state quantum computer by
implementing the Deutsch-Jozsa algorithm in a system of Josephson
charge qubits. In this paper, we propose a scheme to implement
Grover search algorithm with Josephson charge qubits. The scheme is
simple and easily manipulated because any two-charge-qubit can be
selectively and effectively coupled by a common inductance. More
manipulations can be carried out before decoherence sets in. Our
scheme can be realized within the current technology.

\begin{figure}[tbp]
\includegraphics[scale=0.95,angle=0]{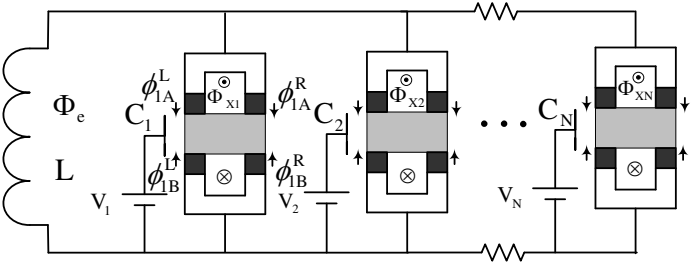}
\caption{ Josephson charge-qubit structure. Each CBP is configured
in the charging regime $E_{ck}\gg E^0_{Jk}$ at low
temperatures $k_BT\ll E_{ck}$. Furthermore, the superconducting gap
$\Delta$ is larger than $E_{ck}$ so that quasiparticle tunneling is
suppressed in the system. } \label{fig1}
\end{figure}

Since the earliest Josephson charge qubit scheme \cite{21-1} was
proposed, a series of  improved schemes \cite{21-2,You2002} have
been brought forward.  Here, we concern the  architecture in Ref.
\cite{You2002}, which is an efficient scalable quantum computing
(QC) architecture via Josephson charge qubits. The Josephson charge
qubits structure is shown in Fig. \ref{fig1}. It consists of
\emph{N} cooper-pair boxes (CPBs) coupled by a common
superconducting inductance L. For the \emph{k}th  cooper-pair box, a
superconducting island with charge $Q_{k}=2en_{k}$ is weakly coupled
by two symmetric direct current superconducting quantum interference
devices (dc SQUIDs) biased by an applied voltage through a gate
capacitance $C_{k}$. Assume that the two symmetric dc SQUIDs are
identical and all Josephson junctions have Josephson coupling energy
$E_{Jk}^{0}$ and capacitance $C_{Jk}$. The self-inductance effect of
each SQUID loop is usually neglected because of very small size
($~1\mu m$) of the loop. Each SQUID pierced by a magnetic flux
$\Phi_{Xk}$ provides an effective coupling energy
$-E_{Jk}(\Phi_{Xk})\cos\phi_{kA(B)}$ with
$E_{Jk}(\Phi_{Xk})=2E_{Jk}^{0}\cos(\pi\Phi_{Xk}/\Phi_0)$, and the
flux quantum $\Phi_0=h/2e$. The effective phase drop $\phi_{kA(B)}$,
with subscript $A(B)$ labeling the SQUID above (below) the island,
equals the average value, $[\phi_{kA(B)}^L+\phi_{kA(B)}^R]/2$, of
the phase drops across two Josephson junctions in the dc SQUID, with
superscript $L(R)$ denoting the left (right) Josephson junction.

For any given cooper-pair box, say $i$, when
$\Phi_{Xk}=\frac{1}{2}\Phi_0$ and $V_{Xk}=(2n_k+1)e/c_k$ for all
boxes except $k=i$, the inductance $L$ connects only the $i$th
cooper-pair box to form a superconducting loop, as shown in Fig.
2($a$).

In the spin-$\frac{1}{2}$ representation, based on charge states
$|0\rangle=|n_i\rangle$ and $|1\rangle=|n_{i+1}\rangle$, the reduced
Hamiltonian of the system becomes \cite{You2002}
\begin{equation}
\label{1}
H=\varepsilon_{i}(V_{Xi})\sigma_z^{(i)}-\overline{E}_{Ji}(\Phi_{Xi},
\Phi_e, L)\sigma_x^{(i)},
\end{equation}
where $\varepsilon_{i}(V_{Xi})$ is controlled by gate voltage
$V_{Xi}$, while the intrabit coupling $\overline{E}_{Ji}(\Phi_{Xi},
\Phi_e, L)$ depends on inductance $L$, applied external flux
$\Phi_e$ through the common inductance, and local flux
$\Phi_{Xi}$  through the two SQUID loops of the $\emph{i}$th
cooper-pair box. By controlling $\Phi_{Xi}$ and $V_{Xi}$, the
operations of Pauli matrice $\sigma_z^{(i)}$ and $\sigma_x^{(i)}$
are achieved. Thus, any single-qubit operations are realized by
Eq. (\ref{1}).
\begin{figure}
\includegraphics[scale=0.95,angle=0]{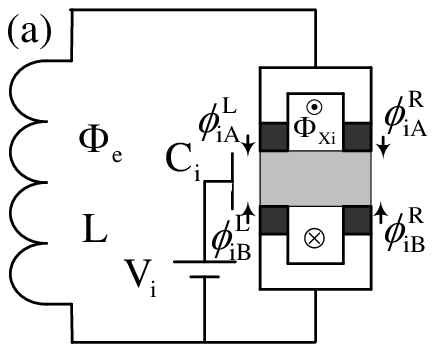}
\includegraphics[scale=0.95,angle=0]{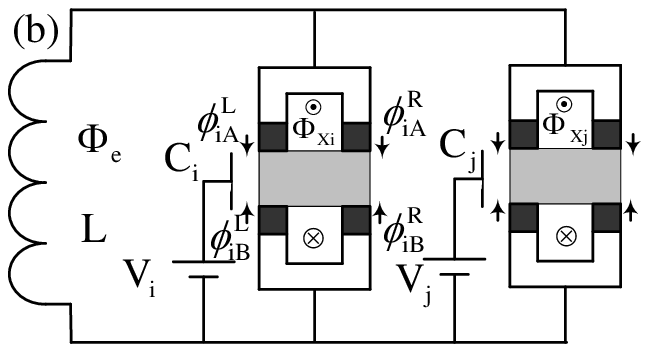}
\caption{(a) single-qubit structure where a CPB is only
 connected to the inductance. (b) Two-qubit structure where two CPBs are connected to the common inductance.} \label{fig2}
\end{figure}

To manipulate any two-qubit, says $i$ and $j$, when
$\Phi_{Xk}=\frac{1}{2}\Phi_0$ and $V_{Xk}=(2n_k+1)e/c_k$ for all
boxes except $k=i$ and $j$, the inductance $L$ is only shared by the
cooper-pair boxes $i$ and $j$ to form superconducting loops, as
shown in Fig. 2($b$), the Hamiltonian of the system can be reduced
to \cite{You2002, You2001}
\begin{equation}
\label{2}
H=\sum_{k=i,j}[\varepsilon_{k}(V_{Xk})\sigma_z^{(k)}-\overline{E}_{Jk}\sigma_x^{(k)}]+\Pi_{ij}\sigma_x^{(i)}\sigma_x^{(j)},
\end{equation}
where the interbit coupling $\Pi_{ij}$ depends on external flux
$\Phi_e$ through the inductance $L$, local fluxes $\Phi_{Xi}$ and
$\Phi_{Xj}$ through the SQUID loops. If letting
$V_{Xk}=(2n_k+1)e/c_k$, Eq.(\ref{2}) can be further reduced to
\begin{equation}
\label{3}
H=-\overline{E}_{Ji}\sigma_x^{(i)}-\overline{E}_{Jj}\sigma_x^{(j)}+\Pi_{ij}\sigma_x^{(i)}\sigma_x^{(j)}.
\end{equation}
For the simplicity of calculation, we set
$\overline{E}_{Ji}=\overline{E}_{Jj}=\Pi_{ij}=\frac{-\pi\hbar}{4\tau}$($\tau$
is a given period of time) by suitably adjusting parameters. Thus,
Eq.(\ref{3}) becomes
\begin{equation}
\label{4}
H=\frac{-\pi\hbar}{4\tau}(-\sigma_x^{(i)}-\sigma_x^{(j)}+\sigma_x^{(i)}\sigma_x^{(j)}).
\end{equation}
According to the Hamiltonian $H$ of Eq. (\ref{4}) above, on the basis
$\{|+\rangle=\frac{1}{\sqrt{2}}(|0\rangle+|1\rangle),
|-\rangle=\frac{1}{\sqrt{2}}(|0\rangle-|1\rangle)\}$, we can obtain
following evolutions:
\begin{subequations}
\label{5}
\begin{equation}
|++\rangle_{ij}\rightarrow e^{-i\pi t/4\tau}|++\rangle_{ij},
\end{equation}
\begin{equation}
|+-\rangle_{ij}\rightarrow e^{-i\pi t/4\tau}|+-\rangle_{ij},
\end{equation}
\begin{equation}
|-+\rangle_{ij}\rightarrow e^{-i\pi t/4\tau}|-+\rangle_{ij},
\end{equation}
\begin{equation}
|--\rangle_{ij}\rightarrow e^{i3\pi t/4\tau}|--\rangle_{ij}.
\end{equation}
\end{subequations}
If we select the optimal interaction time $t=\tau$, and perform a
single-qubit operation $U=e^{i\pi /4}$, we can obtain
\begin{subequations}
\label{6}
\begin{eqnarray}|++\rangle_{ij}\rightarrow |++\rangle_{ij},
\end{eqnarray}
\begin{equation}
|+-\rangle_{ij}\rightarrow |+-\rangle_{ij},
\end{equation}
\begin{equation}
|-+\rangle_{ij}\rightarrow |-+\rangle_{ij},
\end{equation}
\begin{equation}
|--\rangle_{ij}\rightarrow -|--\rangle_{ij}.
\end{equation}
\end{subequations}
From Eqs. (\ref{6}), we can see that the operation of a controlled
phase transform on arbitrary two-qubit has been actually carried
out. On the basis of controlled phase gate, the controlled-not gate
is also obtained by performing a series of single-qubit operations
\cite{You2002}. A sequence of conditional two qubits gates and
single qubit operations constitute a universal element for QC
\cite{LLOYD}, so arbitrary multi-qubit operations can be performed
in our scheme.

\begin{figure}[tbp]
\includegraphics[scale=0.62, angle=0]{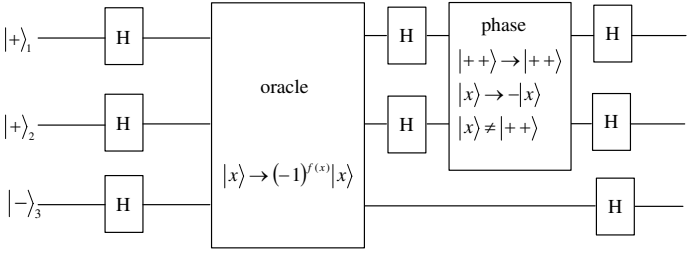}
\caption{The circuit diagram for the two-data-qubit Grover search
algorithm. $H$ denotes Hadamard transformation. $|+\rangle_{1}$ and
$|+\rangle_{2}$ are two data qubits, $|-\rangle_{3}$ is an auxiliary
working qubit.} \label{fig3}
\end{figure}

In the following we only discuss the case of two data qubits, where
we can search a finite item from four items. The circuit diagram of
the Grover search with two data qubits and one auxiliary working
qubit is shown in Fig. {\ref {fig3}}.

Firstly, in order to implement two-data-qubit Grover search
algorithm, we prepare charge qubit 1, 2 and 3 in the state
\begin{equation}
\label{7} |\phi\rangle_{123}=|++-\rangle_{123}.
\end{equation}
By choosing appropriate $\Phi_{Xk}$ and $V_{Xk}$, we can perform
Hadamard operation on charge qubit 1, 2 and 3
\begin{subequations}
\label{8}
\begin{equation}
 |+\rangle_{1} \rightarrow
\frac{1}{\sqrt{2}}(|+\rangle_{1}+|-\rangle_{1}),
\end{equation}
\begin{equation}
|+\rangle_{2} \rightarrow
\frac{1}{\sqrt{2}}(|+\rangle_{2}+|-\rangle_{2}),
\end{equation}
\begin{equation}
|-\rangle_{3} \rightarrow
\frac{1}{\sqrt{2}}(|+\rangle_{3}-|-\rangle_{3}),
\end{equation}
\end{subequations}
So the total state of the charge qubit 1, 2 and 3 becomes
\begin{equation}
\label{9} |\phi\rangle_{123}=\frac{1}{2\sqrt{2}}(|++\rangle_{12}+
|+-\rangle_{12}+|-+\rangle_{12}+|--\rangle_{12})(|+\rangle_{3}-|-\rangle_{3}).
\end{equation}
Obviously, the four items ( $|++\rangle_{12}$ , $|+-\rangle_{12}$ ,
$|-+\rangle_{12}$ , $|--\rangle_{12}$ ) have been stored in the data
qubits before applying the oracle. Without loss of generality, we
search the state $|+-\rangle_{12}$ from the four states. For the
two-data-qubit Grover search algorithm, the oracle has only effect
on the states to be searched. The auxiliary working qubit can be
discarded at this point.

Secondly, in order to achieve the next step in which the function of
the oracle is implemented, we can perform the controlled phase
operation on the two charge qubits as in Eq. (\ref{6}), then perform
single-qubit phase shift on the second qubit. These lead the state
of charge qubit 1 and 2 to

\begin{equation}
\label{10} |\phi\rangle_{12}=\frac{1}{2}(|++\rangle_{12}-
|+-\rangle_{12}+|-+\rangle_{12}+|--\rangle_{12}).
\end{equation}
We perform Hadamard operations on charge qubit 1 and 2 as in Eq.
(\ref{8}$a$) and Eq. (\ref{8}$b$ ). Thus, Eq. (\ref{10}) can be
rewritten as
\begin{equation}
\label{11} |\phi\rangle_{12}=\frac{1}{2}(|++\rangle_{12}+
|+-\rangle_{12}-|-+\rangle_{12}+|--\rangle_{12}).
\end{equation}

Thirdly, for achieving phase change we can perform single-qubit NOT
operations on charge qubits 1, 2, and the controlled phase shift on
the two charge qubits as in Eq. (\ref{6}), which can lead Eq.
(\ref{11}) to
\begin{equation}
\label{12} |\phi\rangle_{12}=\frac{1}{2}(|++\rangle_{12}-
|+-\rangle_{12}+|-+\rangle_{12}-|--\rangle_{12}).
\end{equation}
Then we perform Hadamard operations on charge qubit 1 and 2 as in
Eq. (\ref{8}$a$) and Eq. (\ref{8}$b$ ) again. We can obtain the
state of qubit 1 and 2
\begin{equation} \label{13}
|\phi\rangle_{12}= |+-\rangle_{12}.
\end{equation}

Finally, having been measured by detectors, the state of  charge
qubit 1 and 2 is the result we want to search. The scheme can
generalize to implement a multi-qubit quantum search algorithm by
performing arbitrary multi-qubit operations.

Next, we briefly discuss experimental feasibility of the current
scheme. For the charge qubits in our scheme, the typically
experimental switching time $\tau_{1}$ during a single-qubit
operation is about $0.1ns$ \cite{You2002}. The inductance $L$
$\sim30nH$ in our proposal is experimentally accessible. In the
earlier design \cite{21-2}, the inductance $L$ is about $3.6\mu H$,
which is difficult to make at nanometer scales. Another improved
design \cite{Makhlin} greatly reduces the inductance to $\sim120nH$,
which is about 4 times larger than the one used in our scheme. The
fluctuations of voltage sources and fluxes result in decoherence for
all charge qubits. The gate voltage fluctuations play a dominant
role in producing decoherence. The estimated dephasing time is
$\tau_4\sim10^{-4 }s$ \cite{Makhlin}, which is allowed in principle
$10^6$ coherent single-qubit manipulations. Owing to using the probe
junction, the phase coherence time is only about $2ns$
\cite{Nakamura2,Nakamura2002}. In this setup, background charge
fluctuations and the probe-junction measurement may be two major
factors in producting decoherence \cite{You2002}. The charge
fluctuations are principal only in low-frequency region and can be
reduced by the echo technique \cite{Nakamura2002} and by controlling
the gate voltage to the degeneracy point, but an effective technique
for suppressing charge fluctuations is highly desired.

It is necessary to give the following explains. Our scheme is a
perfect one in which the fidelity and successful probability of
final result are both 1.0. The current technology can realize our
scheme. If experimental technology is imperfect, the SQUIDs will be
asymmetric and the local flux $\Phi_e$ will not completely suppress
the Josephson coupling in the considered circuit. These can result
in some residual Josephson coupling except the required coupling. In
this case, the fidelity and successful probability of the final
result will decrease slightly, but the arbitrary single-qubit and
two-qubit operations can be still performed.

In a summary, we have proposed a new scheme to implement the Grover
search algorithm with Josephson charge qubits. Our scheme is simple
but fairly efficient, and easily manipulated because any
two-charge-qubit can be selectively and effectively coupled by a
common inductance. More manipulations can be carried out before
decoherence sets in. Our scheme can be realized within the current
technology. The implementation of the algorithm would be a key step
to scale more complex quantum algorithms and very important for
constructing a real quantum computer with Josephson charge qubits.

\begin{acknowledgments}
This work was supported partially by Natural Science Foundation of
China under Grants No. 60678022 and No. 10674001, the Doctoral Fund
of Ministry of Education of China under Grant No. 20060357008, the
Key Program of the Education Department of Anhui Province under
Grant No. 2006KJ070A, the Program of the Education Department of
Anhui Province under Grant No. 2006KJ057B and the Talent Foundation
of Anhui University, Anhui Key Laboratory of Information Materials
and Devices (Anhui University).

\end{acknowledgments}

\end{document}